\documentclass[epj,nopacs]{svjour}
\usepackage{epsfig}
\usepackage{latexsym}
\usepackage{amssymb}

 \newcommand{\be}{\begin{equation}}
 \newcommand{\ee}{\end{equation}}
 \newcommand{\bea}{\begin{eqnarray}}
 \newcommand{\eea}{\end{eqnarray}}
 \newcommand{\nn}{\nonumber}

 \journalname{EPJ C}
 
\begin{document}

\setcounter{page}{1}
\title{
Complete description of polarization effects in
\boldmath{$e^+e^-$} pair production by a photon in the field of a
strong laser wave}
\titlerunning{Polarization effects in $e^+e^-$ pair
production by a photon in the field of a strong laser wave}
\authorrunning{D.Yu.~Ivanov et al.}
\author{D.Yu.~Ivanov \inst{1,}\thanks{e-mail: d-ivanov@math.nsc.ru} %
\and G.L.~Kotkin \inst{2,}\thanks{e-mail: kotkin@math.nsc.ru} %
\and V.G.~Serbo \inst{2,}\thanks{e-mail: serbo@math.nsc.ru}}
\institute{Sobolev Institute of Mathematics, Novosibirsk, 6300090,
Russia \and Novosibirsk State University, Novosibirsk, 630090
Russia}

\date{Received: November 20, 2004}


\abstract{ We consider production  of a $e^+e^-$ pair by a
high-energy photon in the field of a strong laser wave. A
probability of this process for circularly or linearly polarized
laser photons and for arbitrary polarization of all other
particles is calculated. We obtain the complete set of functions
which describe such a probability in a compact invariant form.
Besides, we discuss in some detail the polarization effects in the
kinematics relevant to the problem of $e\to \gamma$ conversion at
$\gamma \gamma$ and $\gamma e$ colliders. }

 \maketitle


\section{Introduction}

The $e^+e^-$ pair production by two photons,
 \be
  \gamma(k_1) +\gamma (k_2) \to e^+(p_+) +e^- (p_-),
  \label{1}
 \ee
was calculated by Breit and Wheeler in 1934 (see, for example, the
text-book~\cite{BLP}, \S 88). The polarization properties of this
process were considered in~\cite{McMaster-61,Khoze-70,BG-02}. With
the growth of the laser field intensity, a high-energy photon
starts to interact coherently with $n$ laser photons,
 \be
\gamma(k_1) +n\,\gamma_{\rm L} (k_2) \to e^+(q_+) +e^- (q_-)  \,,
 \label{2}
 \ee
thus the  Breit-Wheeler process becomes non-linear. Such a process
with absorption of $n=4\div 5$ linearly polarized laser photons
was observed in the recent experiment at SLAC~\cite{SLAC}. The
polarization properties of the process (\ref{2}) are especially
important for future $\gamma \gamma$ and $\gamma e$ colliders
where this process can be a significant background for the laser
conversion of $e\to \gamma$ in the conversion region
(see~\cite{GKST-83}, \cite{GKPnQED} and the literature therein).
In this case the non-linear Breit-Wheeler process must be taken
into account in simulations of the processes in a conversion
region. For comprehensive simulation, including processes of pair
production and multiple electron scattering, one has to know not
only the differential cross section of the process (\ref{2}) with
a given number of the absorbed laser photons $n$, but energy,
angles and polarization of final positrons and electrons as well.
Some particular polarization effects for this process were
considered in~\cite{NR,NNR,R-review,Tsai,Yokoya,GS} and have
already been included in the existing simulation
codes~\cite{Yokoya,Telnov-Code}.

In the present paper we give the complete description of the
non-linear Breit-Wheeler process for the case of circularly or
linearly polarized laser photons and arbitrary polarization of all
other particles. For this purpose we exploit intensively the
results of our recent paper~\cite{IKS-04} devoted to the
cross-channel process --- the non-linear Compton
scattering\footnote{Below we shall quote formulae from this paper
by a double numbering, for example, Eq.~(1.21) means Eq.~(21) from
Ref.~\cite{IKS-04}.}
 \be
   e(q) +n\,\gamma_{\rm L} (k) \to e(q') +\gamma (k')\,.
 \label{3}
 \ee
Let $e$ ($e^{\prime *}$) be the polarization 4-vector of the
initial (final) photon and  $u_{\bf p}$ ($\bar{u}_{{\bf p}'}$) be
the bispinor of the initial (final) electron in the scattering
amplitude of the process (\ref{3}) while $e_1$ ($e_2$) be the
polarization 4-vector  of the initial high-energy (laser) photon
and $v_{{\bf p}_+}$ ($\bar{u}_{{\bf p}_-}$) be the bispinor of the
final positron (electron) in the scattering amplitude of the
process (\ref{2}). Then, basically, the results for the discussed
process can be obtain from the corresponding results for the
non-linear Compton scattering~\cite{IKS-04} by replacements
 \bea
&&q\to -q_+\,,\;\; k\to k_2\,,\;\; q'\to q_-\,,\;\;
  k' \to -k_1\,,
 \nn\\
&&p\to -p_+\,,\;\;p'\to p_-\,,
 \label{4}
  \eea
for the 4-momenta of particles involved and
 \bea
&&e(k) \to e_2 (k_2)\,,\;\; e^{\prime *}(k^{\prime}) \to e_1
(k_1)\,,
 \nn\\
 &&u_{\bf p} \to v_{{\bf p}_+}\,,\;\;\bar{u}_{{\bf p}'}
\to \bar{u}_{{\bf p}_-}
 \label{5}
 \eea
for the amplitudes of their wave functions.

In the next section we describe in detail the kinematics of the
process ({\ref{2}). The effective differential cross section is
obtained in Sect. 3 in the compact invariant form, including the
polarization of all particles. In Sect. 4 we consider the process
(\ref{2}) in the reference frame relevant for the conversion
region at the $\gamma \gamma$ and $\gamma e$ colliders. The
limiting cases are discussed in Sect. 5. In Sect. 6 we present
some numerical results obtained for the range of parameters close
to those in the existing TESLA project~\cite{TESLA}. In the last
section we summarize our results and compare them with those known
in the literature. In Appendix we consider the limit of weak laser
field and present in a compact form cross section of the process
(\ref{1}) with all four particles polarized.

We use the system of units in which the velocity of light $c=1$
and the Plank constant $\hbar=1$. In what follows, we will often
consider the non-linear Breit-Wheeler process in the frame of
reference in which a high-energy photon performs a head-on
collision with laser photons, i.e. in which ${\bf k}_1 \,\parallel
\,(- {\bf k}_2)$. We call this the ``collider system''. In this
frame of reference we choose the $z$-axis along the high-energy
photon momentum ${\bf k}_1$. All azimuthal angles are defined with
respect to one fixed $x$-axis: $\varphi_{\pm}$ of the final
leptons, $\beta_{\pm}$ of its polarization vectors
$\mbox{\boldmath$\zeta$}_{\pm}$ and $\gamma_1$ ($\gamma_2$) of the
direction of the high-energy  photon (laser photon) linear
polarization.

\section{Kinematics}

\subsection{Invariant variables}

The invariant parameter describing the intensity of the laser
field (the parameter of non-linearity) is defined via the mean
value of squared 4-potential $A_\mu(x)$:
 \be
\xi^2=-{e^2\over m^2} \,\langle A_\mu (x) A^\mu(x) \rangle\,,
 \label{6}
 \ee
where $e$ and $m$ are the electron charge and the mass. We use
this definition of $\xi^2$ both for the circularly and linearly
polarized laser photons. Another useful expressions for this
parameter are given by equations (1.4)--(1.5). When describing the
non-linear process (\ref{2}), one has to take into account that in
a laser wave 4-momenta $p_\pm$ of the free $e^{\pm}$ are replaced
by the 4-quasi-momenta $q_\pm$:
 \bea
q_{\pm} &=&p_{\pm}+\xi^2{ m^2\over 2k_2 p_{\pm}}\,k_2\,,
 \label{7}\\
 q^2_{\pm}&=& (1+\xi^2 )\,m^2\equiv m_*^2\,.
 \nn
 \eea
In particular, the energy of the free positron or electron
$E_{\pm}$ is replaced by the quasi-energy
 \be
(q_{\pm})_0=E_{\pm}+\xi^2 {m^2\over 2k_2 p_{\pm}}\,\omega_2\,.
 \label{8}
 \ee

As a result, we deal with the reaction (\ref{2}) for which the
conservation law reads
 \be
 k_1+n\,k_2=q_+ +q_-\,.
 \label{9}
 \ee
Note, that  $k_2 q_{\pm}=k_2p_{\pm}$. It is convenient to use the
following invariant variables:
 \be
x={2k_1 k_2\over m^2}\,, \;\;\; y_\pm\; = {k_2 p_\pm\over
k_1k_2}\,, \;\; y_+ +y_- =1\,.
 \label{10}
 \ee
The threshold $x_n$ of the process (\ref{2}) is determined from
the equality
 \be
(k_1+n\,k_2)^2= m^2\,x_n\, n = (2 m_*)^2\,,\;\; x_n
={4(1+\xi^2)\over n}\,,
 \label{11}
 \ee
therefore, when $x< x_1=4(1+\xi^2)$, the process is possible only
if the number of the absorbed laser photon $n$ is larger then the
threshold value
 \be
n_{\rm th}=  {4(1+\xi^2)\over x}\,.
 \label{12}
 \ee

Further, we introduce the auxiliary combinations
\begin{equation}
s_n=2\sqrt{r_n(1-r_n)},\;\; c_n= 1-2r_n\,,\;\; s_n^2+c_n^2=1\,,
  \label{13}
\end{equation}
where
 \be
 r_n={1+\xi^2\over nx\,y_+y_-}\,.
 \label{14}
 \ee
It is useful to note that these invariants have a simple notion in
the collider frame of reference, namely
 \be
r_n={m_*^2 \over m_\perp^2}\,, \;\;  s_n= {2 m_*|{\bf
p}_\perp|\over m_\perp^2}\,,\;\; c_n= {{\bf p}^2_\perp -
m^2_*\over m_\perp^2} \,,
 \label{15}
 \ee
where
 \be
m_\perp^2=m^2_*+{\bf p}^2_\perp= 2{(k_1q_+) \,(k_1q_-)\over n\,
(k_1k_2)}
 \label{16}
 \ee
and  ${\bf p}_\perp$ is the transverse momentum of the positron in
this system:
 \be
{\bf p}_\perp\equiv \left({\bf p}_{+}\right)_\perp=-\left({\bf
p}_{-}\right)_\perp \,.
 \label{17}
 \ee
Therefore,
 \be
0 \leq s_n < 1\,;\;\;\; 0 < r_n \leq 1\,;\;\; -1 \leq c_n < 1\,.
 \label{18}
 \ee

The minimum and maximum values of the variable $y_\pm$ for the
reaction (\ref{2}) are
 \bea
 \max\{ y_\pm\}&=& y_n\;= {1\over 2}\left(1+ v_n\right)\,,\;\;
v_n= \sqrt{1- {4(1+\xi^2)\over nx}}
 \nn\\
 \min\{ y_\pm\}&=&1- y_n\;= {1\over 2}\left(1- v_n\right)
 \label{19}
 \eea
(here $v_n$ is the quasi-velocity of $e^{\pm}$ in the
center-of-mass system). The value of $y_n$ is close to $1$ for
large $n$, but for a given $n$ it decreases with the growth of the
non-linearity parameter $\xi^2$. With this notation one can
rewrite $r_n$ and $s_n$ in the form
 \bea
 r_n&=&{y_n(1-y_n)\over y_+ y_-}\,,
  \nn\\
  s_n&=& 2\,\sqrt{y_n(1-y_n)}\;{\sqrt{(y_n-y_+)(y_n-y_-)}\over y_+y_-} \,,
   \label{20}
 \eea
from which it follows that
 \be
 s_n \to 0 \;\;\mbox{ at } \;\;y_\pm \to y_n \;\; \mbox{or at} \;\;
 y_\pm \to 1-y_n\,.
 \label{21}
 \ee

It is useful to note that the invariants $x,\; y$  defined in
(1.9) are connected with the corresponding invariants introduced
in the present paper (\ref{10}) by the replacements (\ref{4})
under which we have
 \be
 x\to -xy_+\,,\;\; y\to \frac{1}{y_+}\,.
\label{22}
 \ee
Besides, the combinations of these invariants $s_n,\;c_n,\;r_n$
defined in (1.10)--(1.11) are replaced by the corresponding
combinations (\ref{13})--(\ref{14}) by the rules:
 \be
s_n\to s_n\,,\;\;c_n\to c_n\,,\;\; r_n\to r_n\,.
 \label{23}
 \ee

\subsection{Invariant polarization parameters}

The invariant description of the polarization properties of both
the high-energy and laser photons can be performed in the same way
as in paper~\cite{IKS-04}. We define a pair of unit
4-vectors\footnote{In this definition we take into account that at
the replacements (\ref{4}) we have for vectors (1.19) the
following replacements: $N_\mu \to N_\mu,\; P_\mu \to P_\mu$,
$\sqrt{-P^2}\to \sqrt{-P^2}$, but $\sqrt{-N^2}=m^2 nxy
\sqrt{-P^2}\to -m^2 nx \sqrt{-P^2}= -\sqrt{-N^2}$.}
\begin{equation}
e^{(1)}={N\over-\sqrt{-N^2}}\,,\;\;\;
e^{(2)}={P\over\sqrt{-P^2}}\,,
 \label{24}
\end{equation}
where\
 \bea
N^\mu&=&\varepsilon^{\mu\alpha \beta \gamma} P_\alpha
(-k_1-n\,k_2)_\beta K_\gamma \, ,\;\;\varepsilon^{0123}=+1\,,
 \nn
 \\
 P_\alpha&=&(q_--q_+)_\alpha-{(q_- -q_+)K\over
K^2}\,K_\alpha \,,
 \nn\\
 K_\alpha&=&(n\,k_2-k_1)_\alpha\,,
 \nn\\
\sqrt{-N^2}&=&m^2x\, n\,\sqrt{-P^2}\,,\;\;\;\sqrt{-P^2} =m \,
{s_n\over r_n} \sqrt{1+\xi^2}\,.
 \nn
 \eea
The 4-vectors $e^{(1)}$ and $e^{(2)}$ are orthogonal to each other
and to the 4-vectors $k_1$ and $k_2$:
 \be
e^{(i)} e^{(j)} = - \delta_{ij}, \;\; e^{(i)} k_1 = e^{(i)} k_2
=0; \;\; i,\, j = 1,\,2 \,.
 \label{25}
 \ee
In the collider system they have only transverse component:
 \bea
{\bf e}_\perp^{(1)}&=&{{\bf p}_\perp\times {\bf k}_1\over |{\bf
p}_\perp\times {\bf k}_1|}\,,\;\;\; {\bf e}_\perp^{(2)}=-{{\bf
p}_\perp\over |{\bf p}_\perp|}\,,
 \nn
 \\
e_0^{(j)}&=&e_z^{(j)}=0\,.
 \label{26}
 \eea
Therefore, the vector ${\bf e}^{(2)}$ is in the plane of ${\bf
k}_1$ and ${\bf p}_+$ (the plane of scattering) and ${\bf
e}^{(1)}$ is perpendicular to that plane. Note that the vectors
${\bf e}^{(1)}$, ${\bf e}^{(2)}$ and ${\bf k}_2$ form a
right-handed set as well as the vectors ${\bf e}^{(1)}$, $(-{\bf
e}^{(2)})$ and ${\bf k}_1$.

Let $\xi_j$ be the Stokes parameters for the high-energy photon
which are defined with respect to the 4-vectors $e^{(1)}$ and
$(-e^{(2)})$ and $\tilde{\xi}_j$ be the Stokes parameters for the
laser photon which are defined with respect to the 4-vectors
$e^{(1)}$ and $e^{(2)}$.

These parameters are related to the Stokes parameters which were
used in the description of the non-linear Compton
scattering~\cite{IKS-04}. At the replacements (\ref{5}) the
Sto\-kes parameters $\xi_j$ and $\xi'_j$ for the laser and final
photon in the non-linear Compton scattering are replaced by the
Sto\-kes parameters for the process (\ref{2}) according to the
rules:
 \be
 \xi_j \to \tilde{\xi}_j\,,\;\;\xi_{1,\,3}' \to
\xi_{1,\,3}\,,\;\; \xi_{2}' \to -\xi_{2}\,.
 \label{27}
 \ee
The sign minus in the last rule, $\xi_{2}' \to -\xi_{2}$, is due
to the fact that the polarization matrix of the final photon in
the Compton scattering $\rho^{\prime \mu\nu}$ is replaced at the
replacements (\ref{5}) by the matrix of the initial high-energy
photon $\rho^{\nu\mu}_1$.

As for the polarization of the final positron and electron, it is
necessary to distinguish the polarization vector \mbox{\boldmath
$\zeta$}$_{\pm}^{(f)}$ of the final $e^\pm$ as resulting from the
scattering process itself from the detected polarization
\mbox{\boldmath $\zeta$}$_\pm$ which enters the effective cross
section and which essentially represents the properties of the
detector as selecting one or other polarization of the final
lepton (for detail see~\cite{BLP}, \S 65). The vector {\boldmath
$\zeta$}$_\pm$ also determines the lepton-spin 4-vector
 \be
a_\pm=\left({\mbox {\boldmath $\zeta$}_\pm{\bf p}_\pm\over m}\,,\,
\mbox{\boldmath $\zeta$}_\pm+ {{\bf p}_\pm\,(\mbox{\boldmath
$\zeta$}_\pm {\bf p}_\pm) \over m(E_\pm+m)}\right)
 \label{28}
 \ee
and the mean helicity of the final leptons
\begin{equation}
\langle\lambda_\pm\rangle={\mbox{\boldmath $\zeta$}_\pm{\bf
p}_\pm\over 2|{\bf p}_\pm|}\,.
 \label{29}
\end{equation}

Now we have to define invariants which describe the polarization
properties of the final positrons and electrons. Similar to the
approach in~\cite{IKS-04}, we define the two sets of units
4-vectors:
 \bea
e_1^{\pm}&=&\mp \,e^{(1)}\,,\;\;\; e_2^{\pm}=\pm\, e^{(2)} \mp
{\sqrt{-P^2}\over m^2 xy_\pm}\,k_2\,,
 \nn
 \\
e_3^{\pm}&=&{1\over m}\left(p_\pm-{2\over xy_\pm}\,k_2\right)\,.
 \label{30}
 \eea
These vectors satisfy the conditions:
 \bea
e_i^+ e_j^+&=&-\delta_{ij}\;,\;\;\;e_j^+ p_+=0\;
 \nn\\
 e^{-}_i
e^{-}_j&=& -\delta_{ij}\; , \;\;\; e^{-}_j p_{-}=0\,,
 \label{31}
 \eea
besides, $e^{-}_j\leftrightarrow e_j^+$ under the exchange
electron $\leftrightarrow$ positron.

It allows us to represent the 4-vectors $a_\pm$  in the following
covariant form
\begin{equation}
a_\pm=\sum_{j=1}^3 \zeta_j^{\pm} e_j^{\pm}\,,
 \label{32}
\end{equation}
where
 \be
 \zeta_j^{\pm}=-a_\pm e_j^{\pm}\,.
 \label{33}
\end{equation}
The invariants $\zeta_j^{\pm}$ describe completely the detected
polarization properties of the final $e^{\pm}$. It is not
difficult to find that $\zeta_1^{\pm}$ is the polarization
perpendicular to the scattering plane. Besides, in the frame of
reference, in which the lepton momentum ${\bf p}_\pm$ is
anti-parallel to the laser photon momentum ${\bf k}_2$, the
invariant $\zeta_3^{\pm}=2\langle\lambda_\pm\rangle$. Furthermore,
we will show that for the practically important case, relevant for
the $e \to \gamma$ conversion, this frame of reference almost
coincides with the collider system.

Note, that the invariants $\zeta_j^{\pm}$ are connected with the
invariants $\zeta_j$ and $\zeta'_j$ for the initial and final
electron in the non-linear Compton scattering. The corresponding
rules are:
 \be
\zeta_j \to - \zeta^{+}_j\,,\;\; \zeta_j' \to \zeta^{-}_j \,.
 \label{34}
 \ee
The sign minus in the first rule, $\zeta_j \to - \zeta^{+}_j$, is
due to the fact that at the replacements (\ref{5}) the unit
4-vectors for the initial electron in the Compton scattering $e_j$
is replaced by the unit 4-vectors of the final positron $e_j^+$
with an additional minus sign, $e_j \to - e_j^+$.

To clarify the meaning of invariants $\zeta_j^{\pm}$ , it is
useful to note that
 \begin{equation}
\zeta_j^{\pm} ={\mbox{\boldmath $\zeta$}}_\pm\,{\bf n}_j^{\pm}\,,
 \label{35}
\end{equation}
where the corresponding 3-vectors are
\begin{equation}
{\bf n}_j^{\pm} ={\bf e}_j^{\pm} -{{\bf p}_\pm\over
E_\pm+m}\;e_{j0}^{\pm}
 \label{36}
\end{equation}
with $e_{j0}^{\pm}$ being a time component of the 4-vector
$e_{j}^{\pm}$ defined in (\ref{30}).  Using the properties
(\ref{31}) of the 4-vectors $e_j^{\pm}$, one can check that
\begin{equation}
{\bf n}_i^+ \; {\bf n}_j^+ =\delta_{ij}\,,\;\;{\bf n}_i^- \; {\bf
n}_j^- =\delta_{ij}\,.
 \label{37}
\end{equation}
As a result, the polarization vector
$\mbox{\boldmath$\zeta$}_\pm$ has the form
\begin{equation}
\mbox{\boldmath$\zeta$}_\pm=\sum^3_{j=1} \zeta_j^{\pm}\;{\bf
n}_j^{\pm}\, .
 \label{38}
\end{equation}

\section{Cross section in the invariant form}

\subsection{General relations}

The usual notion of the cross section is not applicable for the
reaction (\ref{2}) and usually its description is given in terms
of the probability of the process per second $\dot{W}^{(n)}$.
However, for the procedure of simulation in the conversion region
as well as for the simple comparison with the linear process
(\ref{1}), it is useful to introduce the ``effective cross
section'' given by the definition
 \be
d\sigma^{(n)}= {d\dot{W}^{(n)}\over j}\,,
 \label{39}
 \ee
where
 \be
j= {(k_1\,k_2) \over \omega_1\,\omega_2}\,n_{\rm L} ={m^2 \,x\over
2 \omega_1 \omega_2} \, n_{\rm L}
 \label{40}
 \ee
is the flux density of colliding particles (and $n_{\rm L}$ is the
density of photons in the laser wave). Contrary to the usual cross
section, this effective cross section does depend on the laser
beam intensity, i.e. on the parameter  $\xi^2$. The total
effective cross section is defined as the sum over harmonics,
corresponding to the reaction (\ref{2}) with a given number $n$ of
the absorbed laser photons\footnote{In this formula and below the
sum is over those $n$ which satisfy the condition $y< y_n$, i.e.
this sum runs from some minimal value $n_{\min}$ up to $n=\infty$,
where $n_{\min}$ is determined by the equation $y_{n_{\min}-1} < y
< y_{n_{\min}}$.}:
  \be
d\sigma=\sum\limits_{n}d\sigma^{(n)}\,.
 \label{41}
 \ee
The effective differential cross section with arbitrary
polarization of all particles can be presented in the following
invariant form:
 \bea
&&d\sigma(\mbox{\boldmath$\xi$},\,
\tilde{\mbox{\boldmath$\xi$}},\,\mbox{\boldmath$\zeta$}_\pm) =
{r_e^2\over 4x}\;\sum_n \bar{F}^{(n)}\;d\Gamma_n\,,
 \label{42}
 \\
&&d\Gamma_n = \delta (k_1+n\,k_2-q_+-q_-)\;{d^3q_+\over
(q_+)_0}{d^3q_-\over (q_-)_0}\,,
 \nn
 \eea
where $r_e=\alpha/m$ is the classical electron radius, and
$\bar{F}$ can be obtained from the corresponding function $F$ for
the non-linear Compton scattering (1.32) by relation\footnote{The
sign minus in this equation is due to the fact that the
polarization matrix of the initial electron  in the Compton
scattering $\left(\hat{p}+m\right)(1-\gamma^5 \hat{a})/2$ is
replaced at the transition (\ref{5}) to the matrix
$\left(-\hat{p}_++m\right)(1-\gamma^5 \hat{a}_+)/2$, which is the
polarization matrix of the final positron times $(-1)$.}
 \be
\bar{F}^{(n)}=- F^{(n)}\,,
 \label{43}
 \ee
and in the function $F^{(n)}$, defined by equations (1.33),
(1.46)--(1.52), (1.71)--(1.76), the following replacements,
discussed in the previous section, have to be done:
 \bea
x&\to& -xy_+\,,\;\; y\to {1\over y_+}\,,
 \nn
 \\
r_n &\to& r_n\,,\;\; s_n \to s_n\,,\;\; c_n\to c_n\,,
 \label{44}
 \\
\xi_j &\to& \tilde{\xi}_j\,,\;\xi_{1,\,3}' \to \xi_{1,\,3}\,,\;
\xi_{2}' \to -\xi_{2}\,,\;  \zeta_j \to - \zeta^{+}_j,\,\;\zeta_j'
\to \zeta^{-}_j \,.
  \nn
 \eea

We present the function $\bar{F}^{(n)}$ in the form:
 \bea
\bar{F}^{(n)}&=&\bar{F}_0^{(n)}+\sum ^3_{j=1}\left(
G_j^{(n)+}\zeta^+_j\; + \;G_j^{(n)-} \zeta^{-}_j\right)
 \nn\\
 &&+ \sum
^3_{i,j=1}\bar{H}_{ij}^{(n)}\,\zeta^{+}_i\,\zeta^{-}_j \,.
 \label{45}
 \eea
Here the function $\bar{F}_0^{(n)}$ describes the total cross
section for a given harmonic $n$, summed over spin states of the
final particles:
 \be
\sigma^{(n)}(\mbox{\boldmath$\xi$},\,
\tilde{\mbox{\boldmath$\xi$}})= {r_e^2\over x}\; \int
\bar{F}_0^{(n)}\,d\Gamma_n \,.
 \label{46}
 \ee
The terms $G_j^{(n)+} \zeta^{+}_j$ and $G_j^{(n)-} \zeta^{-}_j$ in
(\ref{45}) describe the polarization of the final positrons and
the final electrons, respectively. The last terms
$\bar{H}_{ij}^{(n)} \zeta^{+}_i\,\zeta^{-}_j$ stand for the
correlation of the final particles' polarizations.

From (\ref{45}) one can deduce the polarization of the final
positron $\zeta_j^{(f)+}$ and electron $\zeta^{(f)-}_j$ resulting
from the scattering process itself. According to the usual rules
(see~\cite{BLP}, \S 65), we obtain the following expression for
the polarization of the final positron (electron) (summed over
polarization states of the final electron (positron)):
 \bea
\zeta_j^{(f)\pm}&=& {G_j^\pm \over \bar{F}_0}\,,\;\;
\bar{F}_0=\sum_n \bar{F}_0^{(n)}\,,\;\;G_j^\pm=\sum_n
G_j^{(n)\pm}\,;
 \nn
 \\
j&=& 1,\,2,\,3\,,
 \label{47}
 \eea
therefore, its polarization vector is
 \be
\mbox{ \boldmath $\zeta$}^{(f)}_{\pm}= \sum_{j=1}^3 \, {G_j^\pm
\over \bar{F}_0}\, {\bf n}_j^{\pm}\,.
 \label{48}
 \ee
In the similar way, the polarization properties for a given
harmonic $n$ are described by
\begin{equation}
\zeta_{j}^{(n)(f)\pm}= {G_j^{(n)\pm}\over \bar{F}_0^{(n)}}\,.
 \label{49}
\end{equation}

\subsection{The results for the circularly polarized laser photons}

In this subsection we consider the case of 100\% circularly
polarized laser beam with the Stokes parameters
 \be
\tilde{\xi}_1=\tilde{\xi}_3=0,\;\; \tilde{\xi}_2=P_c= \pm 1\,.
 \label{50}
 \ee
In the considered case almost all dependence on the non-linearity
parameter $\xi^2$ accumulates in three functions:
  \bea
f_n& \equiv&
f_n(z_n)=J_{n-1}^{2}(z_n)+J_{n+1}^{2}(z_n)-2J_{n}^{2}(z_n)\,,
 \nn\\
 g_n& \equiv& g_n(z_n)=\frac{4 n^2 J_{n}^2(z_n)}{z_n^2}\,,
 \label{51}\\
h_n&\equiv& h_n(z_n)=J_{n-1}^{2}(z_n)-J_{n+1}^{2}(z_n)\,,
 \nn
 \eea
where $J_n(z)$ is the Bessel function. The functions (\ref{51})
depend on $x$, $y_\pm$ and $\xi^2$ via the single argument
 \be
z_n= {\xi\over \sqrt{1+\xi^2}}\; n\,s_n\,.
 \label{52}
 \ee
For the small value of this argument one has
\be
f_n=g_n=h_n=\frac{(z_n/2)^{2(n-1)}}{[(n-1)!]^2} \;\; \mbox{ at}
\;\; z_n \to 0\,,
 \label{53}
 \ee
in particular,
 \be
 f_1=g_1=h_1=1
\;\; \mbox{ at} \;\; z_1= 0\,.
 \label{54}
 \ee
It is useful to note that this argument is small for small
$\xi^2$, as well as for the minimum or maximum values of $y_\pm$:
 \bea
 z_n &\to& 0 \;\; \mbox{either at} \;\; \xi^2\to 0\,,
 \nn
 \\
&&\mbox{ or at }\;\; y_\pm\to 1-y_n\,,\;\;\mbox{ or at }\;\;
y_\pm\to y_n\,.
  \label{55}
 \eea

The results of our calculations are the following. The function
$\bar{F}_0^{(n)}$, related to the total cross section (\ref{46}),
reads
 \bea
\bar{F}_0^{(n)}&=&\left(u-2\right)\,f_n+ {s_n^2\over 1+\xi^2}\,
g_n -\left(u-2\right)\, c_n h_n P_c\,\xi_2
 \nn\\
 &&-\left[2(f_n-g_n)+ s^2_n
(1+\Delta)\,g_n\right]\, \xi_3\,.
  \label{56}
 \eea
Here and below we use the notations:
 \be
 u={1\over y_+\, y_-}\,,\;\; \Delta= {\xi^2 \over 1+\xi^2}\,.
 \label{57}
 \ee

The polarization of the final positrons $\zeta_j^{(f)+}$ is given
by (\ref{47})--(\ref{49}) with
  \bea
G_1^{(n)+}&=&-{1\over y_-} {s_n \over \sqrt{1+\xi^2}}\, h_n P_c\,
\xi_1\,,
 \label{58}
 \\
G_2^{(n)+}&=& {s_n \over y_+\sqrt{1+\xi^2}} \left(c_n g_n \xi_2 -
h_n\,P_c\right)
  \nn\\
 &&+{1\over y_-} {s_n \over \sqrt{1+\xi^2}}\, h_n P_c\,\xi_3\,,
  \nn\\
 G_3^{(n)+}&=&-(y_+ -y_-)u\, c_n\,h_n\,P_c
  \nn\\
&&+u \left[ (y_+ -y_-)\,f_n+ {s^2_n\,y_-\over 1+\xi^2}\,g_n\right]
\xi_2\,.
  \nn
 \eea

The polarization of the final electrons $\zeta_j^{(f)-}$ is given
by (\ref{47})--(\ref{49}) with
 \bea
G_1^{(n)-}&=& - \frac{s_n}{y_+\sqrt{1+\xi^2}}h_n P_c\,\xi_1\,,
 \label{59}
 \\
G_2^{(n)-}&=&\frac{{s}_n}{y_-\sqrt{1+\xi^2}}\,\left(c_n g_n \xi_2
- h_n\,P_c\right)
 \nn\\
&&+ \frac{ s_n}{y_+\sqrt{1+\xi^2}}\,h_n P_c\, \xi_3\,,
 \nn\\
G_3^{(n)-}&=& (y_+ -y_-)u {c}_n\,h_nP_c
 \nn\\
&&- u\, \left[(y_+ -y_-)f_n-\frac{s_n^2\,y_+}{1+\xi^2}g_n\right]
\xi_2\,.
 \nn
 \eea

Correlations of the electron and positron polarizations are given
by the functions
 \bea
\bar{H}_{11}^{(n)}&=& 2f_n-\frac{{s}_n^2}{1+\xi^2}\,g_n
 - 2 c_n h_n P_c\,\xi_2
 \nn\\
 &&- \left\{(u-2)(f_n-g_n)
 -\left[1-(u-1)\Delta\right]s_n^2g_n\right\}\xi_3,
 \nn\\
\bar{H}_{12}^{(n)}&=& u \left[(y_+ -y_-)(f_n-g_n) + \left(y_+
-y_-\Delta\right) s_n^2\,g_n\right]\,\xi_1,
 \nonumber \\
\bar{H}_{13}^{(n)}&=& -\frac{ c_n s_n}{y_-\sqrt{1+\xi^2}}g_n
\,\xi_1,
 \nn
  \\
 \bar{H}_{21}^{(n)}&=&-u\,\left[(y_+ -y_-)\,(f_n-g_n) - \left(y_- -
y_+\Delta\right)s_n^2\,g_n\right] \xi_1,
  \nn\\
\bar{H}_{22}^{(n)}&=&2f_n-\frac{{s}_n^2}{1+\xi^2}\,g_n -2 c_n h_n
P_c\,\xi_2
 \nn\\
 &&-\left[(u-2)(f_n-g_n) + \left(u-1 - \Delta
\right)\,s_n^2\,g_n\right]\,\xi_3\,,
 \nn\\
\bar{H}_{23}^{(n)}&=&\, \frac{ {s}_n
}{y_+\sqrt{1+\xi^2}}\,\left({c}_ng_n -h_n P_c\,\xi_2\right)
 \label{60}\\
 &&+ \frac{ c_n s_n}{y_-\sqrt{1+\xi^2}}\,g_n\,\xi_3\,,
 \nn\\
\bar{H}_{31}^{(n)}&=& -\frac{c_n
s_n}{y_+\sqrt{1+\xi^2}}g_n\,\xi_1\,,
 \nn\\
\bar{H}_{32}^{(n)}&=&
\frac{{s}_n}{y_-\sqrt{1+\xi^2}}\,\left(c_n\,g_n-h_n
P_c\,\xi_2\right)
 \nn\\
 &&+ \frac{c_n s_n}{y_+\sqrt{1+\xi^2}}\,g_n \,\xi_3\,,
 \nn\\
\bar{H}_{33}^{(n)}&=&-\left(u-2\right)f_n
 + \left(u-1\right)\frac{{s_n^2}}{1+\xi^2}g_n
 \nn\\
 &&+ (u-2)\,c_n h_n P_c\,\xi_2
 \nn\\
 && +\left[2(f_n-g_n)+(1+\Delta)\,s_n^2\,g_n\right]\,\xi_3\,.
 \nn
 \eea

It should be noted that among the presented functions there are
defined relations connected with the symmetry under the exchange
electron $\leftrightarrow$ positron, namely, under the replacement
 \be
 y_+ \leftrightarrow y_-
 \label{61}
 \ee
one has
 \be
G_j^{(n)+} \leftrightarrow G_j^{(n)-}\,,\;\;\bar{H}_{ij}^{(n)}
\leftrightarrow \bar{H}_{ji}^{(n)}\,.
 \label{62}
 \ee

\subsection{The results for the linearly polarized laser photons}

Here we consider the case of 100\% linearly polarized laser beam.
In this case, the electromagnetic laser field is described by the
4-potential
 \bea
A^\mu(x)& =& A^\mu\,\cos{(kx)}\,,\;\; A^\mu= {\sqrt{2}\,m\over
e}\,\xi \;e_{\rm L}^\mu\,,
 \nn
 \\
e_{\rm L} e_{\rm L} &=& -1\,,
 \label{63}
 \eea
where $e^\mu_{\rm L}$ is the unit 4-vector describing the
polarization of the laser photons, which can be expressed in the
covariant form via the unit 4-vectors $e^{(1,2)}$ given in
(\ref{24}) as follows:
 \be
e_{\rm L} = e^{(1)}\, \sin{\varphi} -e^{(2)}\, \cos{\varphi}\,.
 \label{64}
 \ee
The Stokes parameters $\tilde{\xi}_i$ of the laser photon are
defined with respect to the 4-vectors $e^{(1)}$ and $e^{(2)}$ and
are equal to
\begin{equation}
\tilde{\xi}_1=-\sin{2\varphi},\;\;\; \tilde{\xi}_2=0,\;\;\;
\tilde{\xi_3}=-\cos{2\varphi}\,.
 \label{65}
\end{equation}

The invariants
 \be
\sin{\varphi}= -e_{\rm L}\,e^{(1)}\,,\;\;\;\cos{\varphi}=e_{\rm
L}\,e^{(2)}
 \label{66}
 \ee
have a simple notion in the collider system in which the vectors
$e^{(1,2)}$ are given by (\ref{26}). For the problem discussed it
is convenient to choose the $x$-axis of this frame of reference
along the direction of the laser linear polarization, i.e. along
the vector ${\bf e}_{\rm L}$. With such a choice, the quantity
$\varphi$ is the azimuthal angle of the final positron $\varphi_+$
in the collider system (analogously, the azimuthal angle
$\beta_\pm$ of the $e^\pm$ polarization
$\mbox{\boldmath$\zeta$}_\pm$ is also defined with respect to this
$x$-axis).

When calculating the effective cross section, we found that almost
all dependence on the non-linearity parameter $\xi^2$ accumulates
in three functions:
 \bea
\tilde{f}_n&=& 4\left[A_1(n,\,a,\,b)\right]^2 -4A_0(n,\,a,\,b)
A_2(n,\,a,\,b) \,,
 \nn\\
\tilde{g}_n &=& {4n^2\over z_n^2}\, \left[A_0(n,\,a,\,b)
\right]^2\,,
 \label{67}
 \\
 \tilde{h}_n &=& {4n\over a}\,A_0(n,\,a,\,b)\,A_1(n,\,a,\,b)\,,
 \nn
 \eea
where
 \bea
&&A_k(n,\,a,\,b)
 \nn
 \\
&& =\int\limits_{-\pi}^{\pi} \,\cos^k{\psi}\,\exp {\left[{\rm
i}\left(n\psi-a \sin{\psi} + b\sin{2\psi}\right) \right]}
{d\psi\over 2\pi}\,.
 \label{68}
 \eea
The arguments of these functions are
 \be
a=- z_n \; \sqrt{2}\, \cos{\varphi}\,,\;\; b=  {\xi^2\over
2xy_+y_-}\,\,,
 \label{69}
 \ee
where $z_n$ is defined in (\ref{52}). To find the positron
spectrum, one needs also the functions (\ref{67}) averaged over
the azimuthal angle $\varphi$:
 \be
\langle \tilde{f}_n \rangle =  \int_0^{2\pi} \tilde{f}_n\, {d
\varphi \over 2\pi}\,,\;\; \langle \tilde{g}_n \rangle =
\int_0^{2\pi} \tilde{g}_n\, {d \varphi \over 2\pi}\,.
 \label{70}
 \ee

For small values of $\xi^2 \to 0$ one has
 \be
\tilde{f}_n ,\,\tilde{g}_n ,\, \tilde{h}_n \propto
\left({\xi^2}\right)^{n-1} \, ,
 \label{71}
 \ee
in particular, at $\xi^2=0$
 \be
\tilde{f}_1=\langle \tilde{f}_1 \rangle= \langle \tilde{g}_1
\rangle= \tilde{h}_1=1\,,\;\; \tilde{g}_1=1+\cos2\varphi\,.
 \label{72}
 \ee

The results of our calculations are the following. First, we
define the auxiliary functions
 \bea
X_n&=& \tilde{f}_n-(1+c_n)\left[ (1-\Delta\, r_n)\,\tilde{g}_n
-\tilde{h}_n\,\cos{2\varphi}\right] \, ,
 \nn \\
Y_n&=&(1+c_n)\tilde{g}_n-2\tilde{h}_n \cos^2\!{\varphi} \, ,
 \label{73}
\\
V_n&=& \tilde{f}_n\,\cos{2\varphi}
 \nn\\
&& + 2 (1+c_n) \left[(1-\Delta\, r_n)\,\tilde{g}_n -
2\tilde{h}_n\, \cos^2\!{\varphi}\, \right]\,\sin^2\!{\varphi} \,,
 \nn
 \eea
where $c_n$, $r_n$ and $\Delta$ are defined in (\ref{13}),
(\ref{14}) and (\ref{57}), respectively.

The function $\bar{F}_0^{(n)}$, related to the total cross section
(\ref{46}), reads
 \bea
\bar{F}_0^{(n)}&=&(u-2)\,\tilde{f}_n + {s_n^2\over 1+\xi^2}\,
\tilde{g}_n - 2\, X_n\, \, \xi_1\sin{2\varphi}
 \nn
 \\
&& +\left(2\, V_n-\frac{s_n^2}{1+\xi^2}\,\tilde{g}_n\right) \,
\xi_3\,.
 \label{74}
   \eea

The polarization of the final positrons $\zeta_j^{(f)+}$ is given
by (\ref{47})--(\ref{49}) with
  \bea
G_1^{(n)+}&=&- {s_n \over y_+\sqrt{1+\xi^2}}  \tilde{h}_n \,
\xi_2\,
 \sin{2\varphi}\,,
 \label{75}
 \\
G_2^{(n)+}&=&{s_n \over y_+\sqrt{1+\xi^2}}\,Y_n   \, \xi_2\,,
  \nn
  \\
  G_3^{(n)+}&=&u\left[(y_+ -y_-)\,\tilde{f}_n+ {s^2_n\,y_-\over
1+\xi^2}\,\tilde{g}_n\right] \,\xi_2\,.
  \nn
 \eea

The polarization of the final electrons $\zeta_j^{(f)-}$ is given
by (\ref{47})--(\ref{49}) with
 \bea
G_1^{(n)-}&=& -\frac{ s_n}{y_-\sqrt{1+\xi^2}}\, \tilde{h}_n \,
\,\xi_2 \sin{2\varphi}\,,
 \label{76}
 \\
G_2^{(n)-}&=& \frac{ s_n}{y_-\sqrt{1+\xi^2}}\, Y_n \, \xi_2\,,
 \nn\\
G_3^{(n)-}&=&-u\left[ (y_+ -y_-)\tilde{f}_n-
\frac{s_n^2\,y_+}{1+\xi^2}\, \tilde{g}_n \right] \,\xi_2\,.
 \nn
 \eea

Correlations of the electron and positron polarizations are given
by the functions
 \bea
\bar{H}_{11}^{(n)}&=&
\left[2\tilde{f}_n-\frac{{s}_n^2}{1+\xi^2}\,\tilde{g}_n\right]-(u-2)\,
X_n \,\xi_1\,\sin{2\varphi}
 \nn\\
 &&+\left[(u-2) V_n+\frac{s_n^2}{1+\xi^2}\,
\tilde{g}_n\right]\,\xi_3\,,
 \nn\\
\bar{H}_{12}^{(n)}&=& -u\left[ (y_+ -y_-)V_n
-\frac{s_n^2\,y_+}{1+\xi^2}\, \tilde{g}_n \right]\,\xi_1
 \nn\\
 &&- u(y_+ -y_-)\,X_n\,\xi_3\,\sin{2\varphi}\,,
 \nonumber \\
\bar{H}_{13}^{(n)}&=&- {s_n \over y_+\sqrt{1+\xi^2}}  \,
\tilde{h}_n \,\sin{2\varphi}
 \nn\\
&& -{s_n  \over y_- \sqrt{1+\xi^2}}\,\left(Y_n\,\xi_1-
\tilde{h}_n\,\xi_3\,\sin{2\varphi}\right)\,,
 \label{77}
  \\
 \bar{H}_{21}^{(n)}&=&  u\left[ (y_+ -y_-)\, V_n + { s_n^2\,y_-
 \over (1+\xi^2)}\, \tilde{g}_n \right]\,\xi_1
  \nn\\
  &&+ u(y_+ -y_-)\,X_n\,\xi_3\,\sin{2\varphi}\,,
  \nn\\
 \bar{H}_{22}^{(n)}&=&\left[2\tilde{f}_n-\frac{{s}_n^2}
 {1+\xi^2}\,\tilde{g}_n\right] -(u-2) X_n\,\xi_1 \,\sin{2\varphi}
 \nn\\
 && +\left[(u-2)V_n  - (u-1)\frac{s_n^2}{1+\xi^2} \,
\tilde{g}_n\right]\,\xi_3\,,
 \nn\\
\bar{H}_{23}^{(n)}&=&{s_n \over y_+\sqrt{1+\xi^2}}\, Y_n
 \nn\\
 &&+ {s_n  \over y_- \sqrt{1+\xi^2}}\left(\tilde{h}_n \,\xi_1
\sin{2\varphi}+Y_n
 \,\xi_3\right)\,,
 \nn\\
\bar{H}_{31}^{(n)}&=&- {s_n \over y_-  \sqrt{1+\xi^2}}\,
\tilde{h}_n \, \sin{2\varphi}
 \nn\\
 &&-{s_n \over y_+\sqrt{1+\xi^2}} \left(Y_n  \,\xi_1-\tilde{h}_n\,\xi_3\,
\sin{2\varphi}\right)\,,
 \nn\\
\bar{H}_{32}^{(n)}&=&{s_n \over y_- \sqrt{1+\xi^2}}\, Y_n
 \nn \\
 &&+ {s_n \over y_+\sqrt{1+\xi^2}} \left(\tilde{h}_n
\,\xi_1\,\sin{2\varphi}+Y_n\,\xi_3\right)\,,
 \nn\\
\bar{H}_{33}^{(n)}&=&-\left[(u-2) \tilde{f}_n -(u-1)
\frac{{s_n^2}}{1+\xi^2}\tilde{g}_n \right]+ 2 X_n \xi_1
\,\sin{2\varphi}
 \nn\\
 &&  -\left( 2\, V_n - \frac{s_n^2}{1+\xi^2}\,
\tilde{g}_n\right)\,\xi_3\,.
 \nn
 \eea

The presented functions obey the same relations (\ref{62}),
connected with the symmetry (\ref{61}) under the exchange electron
$\leftrightarrow$ positron,  as for the case of the circularly
polarized laser photons.

\section{Going to the collider system}

\subsection{Exact relations}

As an example of the application of the above formulae, let us
consider the non-linear Breit-Wheeler process in the collider
system defined in Sect. 1. In such a frame of reference the
invariants (\ref{10}) and a phase volume element in (\ref{42}) are
equal to
\begin{equation}
x={4\omega_1 \omega_2\over m^2}\,,\;\; y_\pm= {E_\pm
+(p_\pm)_z\over 2\omega_1}\,,\;\; d\Gamma_n=dy_+\;d\varphi_+\,,
 \label{78}
\end{equation}
and the differential cross section, summed over spin states of the
final particles, is
 \be
{d\sigma^{(n)}(\mbox{\boldmath$\xi$},\,\tilde{\mbox{\boldmath$\xi$}})\over
dy_+ \, d\varphi_+} = {r_e^2\over x}\; \bar{F}_0^{(n)} \,.
 \label{79}
 \ee
Integrating this expression over $\varphi_+$ and then over $y_+$,
we obtain
 \bea
{d\sigma^{(n)}(\mbox{\boldmath$\xi$},\,\tilde{\mbox{\boldmath$\xi$}})\over
dy_+}& =& {2\pi r_e^2\over x}\, \langle \bar{F}_0^{(n)} \rangle
\,,
  \label{80}
  \\
\sigma^{(n)}(\mbox{\boldmath$\xi$},\,\tilde{\mbox{\boldmath$\xi$}})
&=&{2\pi r_e^2\over x}\, \int_{1-y_n}^{y_n} \langle
\bar{F}_0^{(n)} \rangle \, dy_+\,.
 \nn
 \eea

To find $\langle \bar{F}_0^{(n)} \rangle$ we have to know the
azimuthal dependence of the Stokes parameters $\xi_i$ for the
high-energy photon. These parameters are defined with respect to
the polarization vectors ${\bf e}_\perp^{(1)}$ and $(-{\bf
e}_\perp^{(2)})$ given by (\ref{26}), i.e. they are defined with
respect to the $x^{\prime}y^{\prime}z^{\prime}$-axes which are
fixed to the scattering plane. The $x'$-axis is along ${\bf
e}_\perp^{(1)}$ and perpendicular to the scattering plane. The
$y'$-axis is directed along  $(-{\bf e}_\perp^{(2)})$ and,
therefore, it is in that plane.  Let $\check{\xi}_i$ be the Stokes
parameters for the high-energy photons, fixed to the $xyz$-axes of
the collider system. These Stokes parameters are connected with
$\xi_i$ by the following relations:
 \bea
\xi_1 &=& -\check{\xi}_1 \cos{2 \varphi_+}+ \check{\xi}_3 \sin{2
\varphi_+}\,, \; \; \xi_2 = \check{\xi} _2 \,,
 \nn
 \\
\xi_3 &=& -\check{\xi}_3 \cos{2 \varphi_+}- \check{\xi}_1 \sin{2
\varphi_+}\,.
 \label{81}
 \eea

Inserting these relations into the effective cross section, we
find
 \bea
\langle \bar{F}_0^{(n)} \rangle&=& (u-2)\,f_n+ {s_n^2\over
1+\xi^2}\, g_n
 \nn
 \\
&&-(u-2)\, c_n \,h_n\,{\xi_2}\,P_c
 \label{82}
 \eea
for the case of the circularly polarized laser photons and
 \bea
\langle \bar{F}_0^{(n)} \rangle&=&(u-2)\,\langle \tilde{f}_n
\rangle + {s_n^2\over 1+\xi^2}\,\langle \tilde{g}_n \rangle
 \nn
 \\
 &&- \left[ 2\langle\tilde{f}_n\rangle-
(2+2c_n-s_n^2\Delta)\,\langle \tilde{g}_n\rangle \right.
 \nn
 \\
&&+\left. (1+c_n)^2\, \langle\tilde{g}_n\cos{2\varphi}\rangle
\right]\,\check{\xi}_3
 \label{83}
 \eea
for the case of the linearly polarized laser
photons.\footnote{Remind that  in the case of linearly polarized
laser photon we choose $x$-axis along the direction of laser
linear polarization vector ${\bf e}_{\rm L}$.} This means that for
the circularly polarized laser photons the differential
$d\sigma^{(n)}(\mbox{\boldmath$\xi$},\,\tilde{\mbox{\boldmath$\xi$}})
/ dy_+$ and the total
$\sigma^{(n)}(\mbox{\boldmath$\xi$},\,\tilde{\mbox{\boldmath$\xi$}})$
cross sections for a given harmonic $n$ do not depend on the
degree of the linear polarization of the high-energy photon while
for the linearly polarized laser photons these cross sections do
not depend on the degree of the circular polarization of the
high-energy photon.

The polarization vector of the final $e^\pm$ is determined by
(\ref{47})--(\ref{49}), but, unfortunately, the unit vectors ${\bf
n}_j^{\pm}$ in this equation have no simple form. Therefore, we
have to find the characteristics that are usually used for
description of the positron (electron) polarization --- the mean
helicity $\langle \lambda_\pm\rangle$ and its transverse (to the
vector ${\bf p}_\pm$) polarization $\mbox{ \boldmath
$\zeta$}^{\pm}_\perp$ in the considered collider system. For this
purpose we introduce unit vectors $\mbox{
\boldmath$\nu$}_j^{\pm}$, one of them is directed along the
momentum of the final lepton ${\bf p}_{\pm}$ and two others are in
the plane transverse to this direction:
 \bea
&&\mbox{\boldmath$\nu$}_1^{\pm} ={{\bf k}_1\times {\bf p}_\pm\over
|{\bf k}_1\times {\bf p}_\pm|}\,,\;\;\;
\mbox{\boldmath$\nu$}_2^{\pm} ={{\bf p}_{\pm}\times
\mbox{\boldmath$\nu$}_1^{\pm} \over|{\bf p}_{\pm}\times
\mbox{\boldmath$\nu$}_1^{\pm}|}, \;\;\;
\mbox{\boldmath$\nu$}_3^{\pm} = {{\bf p}_{\pm}\over |{\bf
p}_{\pm}|}\,;
 \nn
 \\
&&\mbox{\boldmath$\nu$}_i^+ \, \mbox{\boldmath$\nu$}_j^+ =
\delta_{ij}\,,\;\;\mbox{\boldmath$\nu$}_i^- \,
\mbox{\boldmath$\nu$}_j^- = \delta_{ij}\,,\;\;
 \label{84}
 \eea
Therefore,
$\mbox{\boldmath$\zeta$}_{\pm}\,\mbox{\boldmath$\nu$}_1^{\pm}$ is
the transverse polarization of the final $e^{\pm}$ perpendicular
to the scattering plane,
$\mbox{\boldmath$\zeta$}_{\pm}\,\mbox{\boldmath$\nu$}_2^{\pm}$ is
the transverse polarization in that plane and
$\mbox{\boldmath$\zeta$}_{\pm}\,\mbox{\boldmath$\nu$}_3^{\pm}$ is
the doubled mean helicity of the final electron:
 \be
\mbox{\boldmath$\zeta$}_{\pm}\,\mbox{\boldmath$\nu$}_3^{\pm}=
2\langle\lambda_\pm\rangle\,.
 \label{85}
 \ee

Let us discuss the relation between the scalar products
$\mbox{\boldmath$\zeta$}_{\pm}\mbox{\boldmath$\nu$}_j^{\pm}$,
defined above, and the invariants $\zeta_j^{\pm}$, defined in
(\ref{33}), (\ref{35}). In the collider system the vectors
$\mbox{\boldmath$\nu$}_1^{\pm}$ and ${\bf n}_1^{\pm}$ coincide
 \be
\mbox{\boldmath$\nu$}_1^{\pm}={\bf n}_1^{\pm}\,,
 \label{86}
  \ee
therefore,
 \be
\mbox{\boldmath$\zeta$}_{\pm}\,\mbox{\boldmath$\nu$}_1^{\pm} =
\zeta_1^{\pm}\,.
 \label{87}
  \ee
Two other unit vectors ${\bf n}_2^{\pm}$ and ${\bf n}_3^{\pm}$ are
in the scattering plane and they can be obtained from the vectors
$\mbox{\boldmath$\nu$}_2^{\pm}$ and
$\mbox{\boldmath$\nu$}_3^{\pm}$ by the rotation around the axis
$\mbox{\boldmath$\nu$}_1^{\pm}$ on the angle $(-\Delta
\theta_{\pm})$:
 \bea
\mbox{\boldmath$\zeta$}_{\pm}\,\mbox{\boldmath$\nu$}_2^{\pm}&=&
\zeta^{\pm}_2\,\cos{\Delta \theta_{\pm}}+
\zeta^{\pm}_3\,\sin{\Delta \theta_{\pm}}\,,
 \nn
 \\
\mbox{\boldmath$\zeta$}_{\pm}\,\mbox{\boldmath$\nu$}_3^{\pm}&=&
\zeta^{\pm}_3\,\cos{\Delta \theta_{\pm}}-
\zeta^{\pm}_2\,\sin{\Delta \theta_{\pm}}\,,
 \label{88}
 \eea
where
 \be
\cos{\Delta \theta_{\pm}}= \mbox{\boldmath$\nu$}^{\pm}_3{\bf
n}_3^{\pm}\,,\;\; \sin{\Delta \theta_{\pm}}= -
\mbox{\boldmath$\nu$}^{\pm}_3{\bf n}_2^{\pm}\,.
 \label{89}
  \ee
As a result, the polarization vector of the final positron
(electron) (\ref{38}) is expressed as follows:
 \bea
\mbox{\boldmath$\zeta$}^{(f)}_{\pm}&=&
\mbox{\boldmath$\nu$}_1^{\pm}\,{G_1^{\pm}\over \bar{F}_0}\,+
\mbox{\boldmath$\nu$}_2^{\pm} \left( {G_2^{\pm}\over \bar{F}_0}
\cos{\Delta \theta_{\pm}} + {G_3^{\pm}\over \bar{F}_0} \sin{\Delta
\theta_{\pm}} \right)
 \nn
 \\
 &+&
\mbox{\boldmath$\nu$}_3^{\pm} \left( {G_3^{\pm}\over \bar{F}_0}
\cos{\Delta \theta_{\pm}} - {G_2^{\pm}\over \bar{F}_0} \sin{\Delta
\theta_{\pm}} \right)\,.
 \label{90}
 \eea

\subsection{Approximate formulae}

All the above formulae are exact. In  this subsection we give some
approximate formulae useful for application to the important case
of high-energy $\gamma \gamma$ and $\gamma e$ colliders. It is
expected (see, for example, the TESLA project~\cite{TESLA}) that
in the conversion region of these colliders, an electron with the
energy $250$ GeV performs collisions with laser photons having the
energy $\approx 1$ eV per a single photon. The produced
high-energy photons with the energy $\omega_1\leq 200$ GeV can
further collide with the laser photon of energy $\omega_2 \approx
1$ eV, therefore, the invariant $x= 4\omega_1\omega_2/m^2 \leq 4$
corresponds to the case when the $e^+e^-$ pair can be produced
only at $n\geq 2$. Therefore, the most interesting region of
parameter $x$ is around the value $x\approx 4$. The final
electrons and positrons are ulta-relativistic and they are
produced at a small angles $\theta_{\mp}$ with respect to the
$z$-axis (chosen along the momentum ${\bf k}_1$ of the high-energy
photon):
\begin{equation}
\omega_1\gg m\,,\;\; E_\pm \gg m, \;\;\; \theta_\pm \ll 1.
 \label{91}
\end{equation}
In this approximation we have
\begin{equation}
y_\pm\approx{E_\pm \over \omega_1}\,,
 \label{92}
\end{equation}
therefore, (\ref{79}) gives us the distribution of the final
posit\-rons over their energy and the azimuthal angle. Besides,
the lepton polar angle for the reaction (\ref{2}) is
 \be
\theta_\pm \approx {m\sqrt{nx}\over
\omega_1}\;{\sqrt{(y_n-y_\pm)(y_\pm-1+y_n)}\over
 y_\pm}
 \label{93}
 \ee
and $\theta_{\pm} \to 0$ at $y_\pm\to y_n$ or at $y_\pm\to 1-y_n$.
The maximal value of this angle is small
 \bea
\max\{\theta_{\pm}\} &\approx& {2n\,\omega_2\over
m_*}\,\sqrt{1-{4(1+\xi^2)\over nx}}
 \nn\\
&&\mbox{at}\;\; y_\pm={2(1+\xi^2)\over nx}\,.
 \label{94}
 \eea

It is not difficult to check that, for the considered case, the
angle $\Delta \theta_\pm$ between vector
$\mbox{\boldmath$\nu$}_3^{\pm}$ and ${\bf n}^{\pm}_3$ is very
small:
 \bea
\Delta \theta_{\pm} &\approx& |\mbox{\boldmath$\nu$}^{\pm}_{3}
\times {\bf n}^{\pm}_{3}|
 \label{95}
  \\
&\approx& {m \theta_\pm \over 2 E_\pm} \leq {n m\omega_2\over
m_*E_{\pm}}\, \sqrt{1- {4(1+\xi^2)\over nx}}\,.
 \nn
 \eea
This means that the invariants $\zeta_j^{\pm}$ in (\ref{33}),
(\ref{35}) almost coincide with the projections
$\mbox{\boldmath$\zeta$}_{\pm}$ on the vectors
$\mbox{\boldmath$\nu$}_j^{\pm}$, defined in (\ref{84}),
 \bea
\zeta_1^{\pm}&\approx& -\zeta^{\pm}_{\perp}
\sin{(\varphi_{\pm}-\beta_{\pm})}\,,\;\; \zeta^{\pm}_2\approx
-\zeta^{\pm}_{\perp}\cos{(\varphi_{\pm}-\beta_{\pm})}\,,
 \nn
 \\
\zeta^{\pm}_3&\approx& 2\langle\lambda_\pm\rangle\,,
 \label{96}
 \eea
where $\beta_\pm$ is the azimuthal angle of the vector
\mbox{\boldmath$\zeta$}$_\pm$ and $\zeta^\pm_{\perp}$ stands for
the transverse components of the vectors $\mbox{\boldmath
$\zeta$}_\pm$ with respect to the vector ${\bf p}_\pm$. Moreover,
it means that the exact equation (\ref{90}) for the polarization
of the final positron and electron can be replaced with a high
accuracy by the approximate equation
 \be
\mbox{\boldmath$\zeta$}^{(f)}_{\pm} \approx
\sum_{j=1}^3\,{G_j^{\pm}\over \bar{F}_0}\,
\mbox{\boldmath$\nu$}^{\pm}_j \,.
 \label{97}
 \ee

Up to now we deal with the head-on collisions of the laser photon
and the high-energy photon, when the collision angle $\alpha_0$
between the vectors ${\bf k}_1$ and $(-{\bf k}_2)$ was equal zero.
The detailed consideration of the case $\alpha_0\neq 0$ is given
in Appendix B of paper~\cite{IKS-04}. We present here the summary
of such a consideration. If $\alpha_0\neq 0$, the longitudinal
component of the vector ${\bf k}_2$ becomes
$(k_2)_z=-\omega_2\cos{\alpha_0}$ and it appeared to be the
transverse (to the momentum ${\bf k}_1$)  component $({\bf
k}_2)_\perp$. However this transverse component is small, $|({\bf
k}_2)_\perp| \lesssim\omega_2\ll m$, therefore the transverse
momenta of the final particles are almost compensate each other,
$({\bf p}_-)_\perp=({\bf k}_2-{\bf p}_+)_\perp\approx -({\bf
p}_+)_\perp$. The invariants $x$ and $y_\pm$ become (compare with
(\ref{78}), (\ref{92}))
\begin{equation}
x\approx\frac{4\omega_1\omega_2}{m^2}\cos^2{\frac{\alpha_0}{2}},\;\;
y_\pm\approx \frac{E_\pm}{\omega_1}\,.
 \label{98}
\end{equation}
The polarization parameters of the initial photons and final
electrons (positrons) conserve their forms (\ref{50}), (\ref{65})
and (\ref{81}), (\ref{97}). Therefore, the whole dependence on
$\alpha_0$ enters the effective cross section and the
polarizations only via the quantity $x$ (\ref{98}).

\section{Limiting cases}

In this section we consider several limiting cases in which
description of the non-linear Breit-Wheeler process is essentially
simplified.

{\bf (i)} The case of $\;\xi^2 \to 0$. At small $\xi^2$ all
harmonics with $n > 1$ disappear due to the properties (\ref{53}),
(\ref{55}), (\ref{71}), and we have
 \be
d\sigma^{(n)} (\mbox{\boldmath$\xi$},\,
\tilde{\mbox{\boldmath$\xi$}},\,\mbox{\boldmath$\zeta$}_\pm)
 \propto \xi^{2(n-1)}
\;\; \mbox{ at } \;\; \xi^2\to 0\,.
 \label{99}
 \ee
The corresponding expression for $d\sigma^{(1)}$ is given in
Appendix.

{\bf (ii)} The case of $y_\pm\to y_n$ for the circularly polarized
laser photons. In this limit $y_\mp\to 1-y_n$ and
 \be
{\bf p}_\perp \to 0\,,\;\; s_n \to 0\,,\;\; c_n\to -1\,.
 \label{100}
 \ee
In the collider system with $\omega_1 \gg \omega_2$ both leptons
fly along the momentum of the high-energy photon ${\bf k}_1$ with
$(q_\pm)_z \approx (p_\pm)_z > 0$. It means that
 $$
{\bf n}^{+}_3={\bf n}^{-}_3
=\mbox{\boldmath$\nu$}_3^{+}=\mbox{\boldmath$\nu$}_3^{-}\,, \;\;
{\bf n}^{+}_{1,2} =-{\bf n}^{-}_{1,2}
=\mbox{\boldmath$\nu$}_{1,2}^{+}\,,
 $$
therefore, $\zeta_3^\pm$ coincides with the doubled mean $e^\pm$
helicity,
 \be
\zeta_3^\pm = 2\langle \lambda_\pm \rangle\,,\;\;\zeta_1^+
\zeta_1^- + \zeta_2^+ \zeta_2^-
=-(\mbox{\boldmath$\zeta$}_{+})_\perp
\,(\mbox{\boldmath$\zeta$}_{-})_\perp\,.
 \label{101}
 \ee

All harmonics with $n>1$ disappear\footnote{The vanishing of all
harmonics with $n \geq 3$ is due to the conservation of $z$
component of the total angular momentum $J_z$ in this limit: the
initial value of $J_z= \lambda_1-n\,\lambda_2$ can not be equal
for $n \geq 3$ to the final value of $J_z=\lambda_+ +\lambda_-$
(here $\lambda_1$ $(\lambda_2)$ is the helicity of the high-energy
(laser) photon and $\lambda_\pm$ is the helicity of the positron
(electron)). For $n=2$ this argument does not help, since $J_z=\pm
1$ can be realized for the initial and final states. The vanishing
of the second harmonics for $y_\pm\to y_n$ is a specific feature
of the process, related to the facts that the polarization vectors
$e_{1}^{(\lambda_{1})}$ and $e_{2}^{(\lambda_{2})}$ are orthogonal
not only to the 4-vectors $k_1,\, k_2$ but to the 4-vectors
$p_\pm$ as well and that $e_1^{(\lambda_1)}\,
e_1^{(\lambda_1)}=0$.} and only the leptons of the first harmonic
can be produced strictly along the direction of the high-energy
photon momentum:
 \bea
{d\sigma^{(n)} (\mbox{\boldmath$\xi$},\,
\tilde{\mbox{\boldmath$\xi$}},\,\mbox{\boldmath$\zeta$}_\pm) \over
dy_+ d\varphi_+} &=&{r^2_e\over 4x}\, \bar{F}^{(n)}\propto f_n
 \propto (y_n -y_+)^{n-1}
 \nn
  \\
&&\mbox{ at } \;\; y_\pm\to y_n\,,
 \label{102}
 \eea
with
 \bea
{\bar{F}^{(n)}\over f_n}&=& (1+\xi_2 P_c)
 \nn
 \\
 &&\times\left[
{y^2_n+(1-y_n)^2\over y_n(1-y_n)}\, (1-\zeta_3^+
\zeta_3^-)-2(\mbox{\boldmath$\zeta$}_+)_\perp
(\mbox{\boldmath$\zeta$}_-)_\perp)\right]
 \nn\\
 &&\pm {2y_n-1\over y_n(1-y_n)}\,(\xi_2+ P_c)\,
(\zeta_3^+- \zeta_3^-)
  \label{103}
  \eea
(the $\pm$ sign in the last term of (\ref{103}) corresponds to
$y_\pm\to y_n$). In this limit, the final leptons are longitudinal
polarized,
 \bea
\zeta_3^{(n)(f)+}&=&-\zeta_3^{(n)(f)-}=\pm
 {2y_n-1 \over y^2_n+(1-y_n)^2} \,{\xi_2+P_c\over
1+\xi_2 P_c}\,,
 \nn
 \\
\mbox{\boldmath$\zeta$}_\perp^{(n)(f)\pm}&=&0\,.
 \label{104}
 \eea

{\bf (iii)} The case of large $x\gg 1+\xi^2$ for the circularly
polarized laser photons. In this case $1-y_n\approx (1+\xi^2)/(n
x)\ll 1$. The spectrum of the fist harmonic is symmetric under
replacement $y_+ \leftrightarrow 1-y_+$ and has peaks at maximum
or minimum value of $y_+$. The form of these peaks at not too
small transverse momentum of leptons, at $m_*^2 \ll {\bf
p}_\perp^2 \ll m^2 x$, is
 \be
{d\sigma^{(1)}(\mbox{\boldmath$\xi$},\,
\tilde{\mbox{\boldmath$\xi$}} ) \over dy_+} ={2\pi r^2_e\over x}
(1-\xi_2 P_c) \left[{1\over y_+(1-y_+)} -2\right]\,.
  \label{105}
 \ee
The magnitude of these peaks increases with growth of $x$.
Therefore, the regions near this peaks give the dominant
contribution to the cross section for the first harmonic. On the
other hand, all other harmonics vanishes in this regions. As a
result, the cross section for the first harmonic $\sigma^{(1)}$
approximate well the total cross section $\sigma$ summed over all
harmonics. Thus, we consider the regions
 \be
 m_*^2 \ll {\bf p}_\perp^2 \ll m^2 x\,,
\label{106}
 \ee
in which $y_\pm$ is close to $y_1$ or to $1-y_1$ while $y_1$ is
close to unit:
 \be
1-y_1 \approx {1+\xi^2\over x} \ll 1\,, \;\; |s_n| \ll 1 \,, \;\;
c_n \approx +1\,.
 \label{107}
 \ee
In this regions the function $\bar{F}^{(1)}$ becomes large
 \bea
\bar{F}^{(1)}&\approx& {1\over (1-y_+)(1-y_-)}\left[(1-\xi_2 P_c)
(1-\zeta_3^+ \zeta_3^-)\right.
 \nn
 \\
 &&\left. \pm\,(\xi_2- P_c)\,
(\zeta_3^+- \zeta_3^-)\right]
 \label{108}
   \eea
(for $y_\pm$ near $y_1$). Therefore, integrating the effective
cross section for the first harmonic, $d\sigma^{(1)} \propto
1/[(1-y_+)(1-y_-)] $, near its two maximums, i.e. in the regions
 \be
 {1+\xi^2 \over x} \ll 1-y_\pm \ll 1\,,
 \label{109}
  \ee
we find (with logarithmic accuracy) the total cross section
 \bea
&&\sigma(\mbox{\boldmath$\xi$},\,
\tilde{\mbox{\boldmath$\xi$}},\,\mbox{\boldmath$\zeta$}_\pm)
 \nn
 \\
&&= {\pi r_e^2\over x} \, (1-\xi_2 P_c) (1-\zeta_3^+ \zeta_3^-)\,
\ln{x\over 1+\xi^2}\,.
 \label{110}
 \eea
The total cross section, summed over spin states of the final
particles, has the form
 \be
\sigma(\mbox{\boldmath$\xi$},\, \tilde{\mbox{\boldmath$\xi$}}) =
{4\pi r_e^2\over x} \,\left(1-\xi_2 P_c \right)\, \ln{x\over
1+\xi^2}\,.
 \label{111}
  \ee

{\bf (iv)} The case of $y_\pm\to y_n$ for the linearly polarized
laser photons.  In this limit, the argument $a$ of the functions
(\ref{68}) tends to zero, while the argument $b$ tends to the
nonzero constant, $b\to n \xi^2/[2(1+\xi^2)]$; the function
 \be
\tilde{f}_n =\left[J_{(n-1)/2}(b)-J_{(n+1)/2}(b) \right]^2
 \label{112}
 \ee
for odd $n$ and
 \be
\tilde{f}_n ={2\over \xi^2}\left[J_{n/2}(b)\right]^2
 \label{113}
 \ee
for even $n$. The spectra and the polarizations are determined by
the function $\bar{F}^{(n)}$ given for odd harmonics by the
expression
 \bea
{\bar{F}^{(n)}\over \tilde{f}_n}&=& (u_n-2-2\check{\xi}_3)
(1-\zeta_3^+ \zeta_3^-)
 \nn
 \\
 &&-\left[2-(u_n-2)\check{\xi}_3\right](\mbox{\boldmath$\zeta$}_+)_\perp
(\mbox{\boldmath$\zeta$}_-)_\perp
 \label{114}
 \\
 &&\pm (2y_n-1)u_n\,\left[\xi_2\,
(\zeta_3^+ - \zeta_3^-) +\check{\xi_1}(\zeta_1^+\zeta_2^-
-\zeta_2^+\zeta_1^-) \right]
 \nn
  \eea
 and for even harmonics by the expression
 \bea
\bar{F}^{(n)} &=& u_n\,\tilde{f}_n [1+\zeta_3^+ \zeta_3^-
 +\xi_3 (\zeta_1^+\zeta_1^- -\zeta_2^+\zeta_2^-)
 \nn \\
 &&+ \xi_2\,
(\zeta_3^+ + \zeta_3^-)+\xi_1(\zeta_1^+\zeta_2^-
+\zeta_2^+\zeta_1^-) ]
 \label{115}
  \eea
with $u_n = 1/[y_n(1-y_n)]$.

{\bf (v)} The behavior of the cross section near the threshold
 \be
x\to x_n = {4(1+\xi^2)\over n}\,.
 \label{1116}
 \ee
In this limit
 \be
y\to 1/2\,,\;\; {\bf p}_\perp \to 0\,,\;\; s_n \to 0\,,\;\; c_n
\to -1
 \label{117}
 \ee
and the main results can be obtain from the case $y_+\to y_n$ just
by substitution $y_n= 1/2$ into (\ref{102})--(\ref{104}) and
(\ref{112})--(\ref{115}). In particular, for the first harmonic
with the circularly polarized laser photons we have
 \be
\bar{F}^{(1)}= 2(1+\xi_2 P_c)\left(1 - \mbox{\boldmath$\zeta$}_+
\mbox{\boldmath$\zeta$}_-\right)\,.
 \label{118}
   \ee

It is interesting to compare the behavior of the first harmonic
near the threshold and far from the threshold. Let us consider the
pure quantum states for the photons (their helicities
$|\lambda_{1,2}|=1$) and $e^\pm$ (their helicities
$|\lambda_{\pm}|=1/2$) in the collider system with $\omega_1 \gg
\omega_2$. It follows from (\ref{118}) that near the threshold the
photons have the same helicities while the leptons have the
opposite helicities:
 \be
\lambda_{1}=\lambda_{2}\,,\;\;\lambda_{+}=- \lambda_{-}\;\;
 {\rm at }\;\; x\to x_1=4(1+\xi^2)\,.
 \label{119}
   \ee
If $x\gg x_1$, we have to distinguish two regions. In the region
(\ref{100}) of very small transverse momentums of leptons ${\bf
p}_\perp \to 0$, we have the same relations (\ref{119}) [see
(\ref{103})], but in the region (\ref{106}) of relatively large
transverse momentums of leptons the photons have the opposite
helicities as well as the leptons [see (\ref{108})]:
 \bea
\lambda_{1}&=&-\lambda_{2}\,,\;\;\lambda_{+}=- \lambda_{-}
 \label{120}
 \\
&&{\rm at }\;\; x\gg x_1\,,\;\; m^2 x_1 \ll {\bf p}_\perp^2 \ll
m^2 x\,.
  \nn
   \eea

\begin{figure}[!th]
\includegraphics[width=0.47\textwidth]{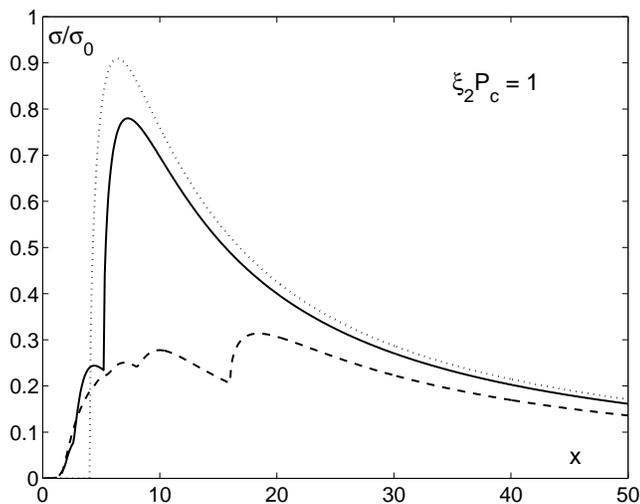}
 \caption{The total effective cross section of the non-linear
Breit-Wheeler process in dependence on $x=4\omega_1\omega_2/m^2$.
The laser and high-energy photons have the same helicity, $P_c
=\xi_2=1$. The dotted curve corresponds to $\xi^2=0$, the solid
curve --- to $\xi^2=0.3$ and the dashed curve
--- to $\xi^2=3$. }
 \label{f1}
\end{figure}

\begin{figure}[!h]
\includegraphics[width=0.47\textwidth]{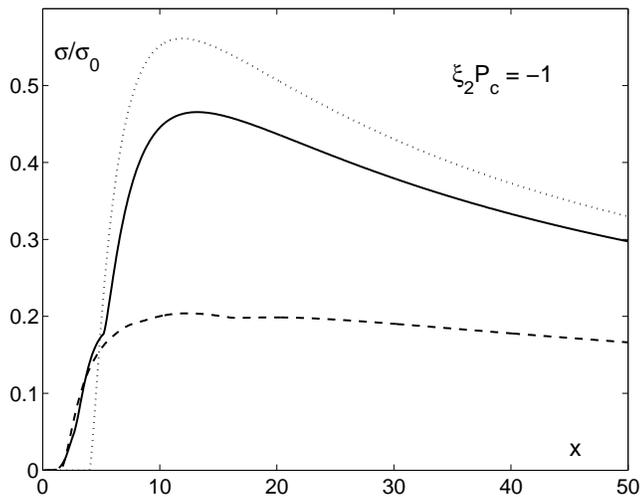}
 \caption{The same as in Fig. 1, but for $P_c =-\xi_2=1$.}
 \label{f2}
\end{figure}

\section{Numerical results related to $\gamma \gamma$ and
$\gamma e$ colliders}

In this section we present some examples which illustrate the
formulae obtained in the previous sections. In figures below we
use notation
 $$
\sigma_0=\pi r_e^2 \approx 2.5\cdot 10^{-25} \;\;\mbox{cm}^2\,.
 $$

The total cross section for the most interesting case of the
circularly polarized laser photons is shown in Figs. \ref{f1} and
\ref{f2} in dependence on the parameter $x= 4\omega_1\omega_2/m^2$
for three values of the non-linearity parameter $\xi^2$ (defined
in (\ref{6})): for $\xi^2=0$ (the linear Breit-Wheeler process
(\ref{1})), for $\xi^2=0.3$ (as in the TESLA project) and larger
by one order of magnitude, $\xi^2=3$. The linear Breit-Wheeler
process has a clear threshold: a pair production is possible only
at $x> x_1=4$. At nonzero value of $\xi^2$, the electron pair
production becomes possible for any $x\geq 0$ due to excitation of
the higher harmonics. Note that, with the increase of $\xi^2$ the
total effective cross section decreases.

It is seen from (\ref{80}) and (\ref{82}) that
all these cross sections depend on the polarization of the initial
photons only via the factor $\xi_2 P_c$. When helicities of the
initial photons are the same ($\xi_2 P_c=1$, see Fig. 1), these
cross sections dominate in the region of not too large $x$
(approximately at $x< 15$), at larger $x$ the cross sections with
the opposite helicities of the initial photons ($\xi_2 P_c=-1$,
see Fig. 2) become dominant.

\begin{figure}[!t]
\includegraphics[width=0.47\textwidth]{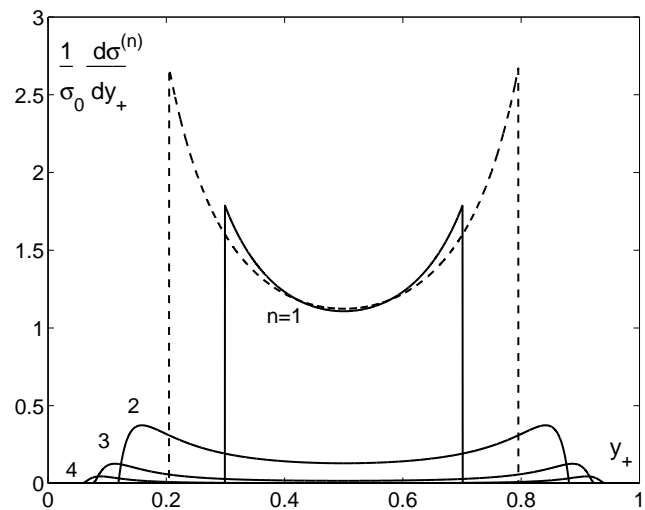}
 \caption{Energy spectra of positrons for different
harmonics $n$ at $x=6.2$ and $\xi^2=0.3$. The laser and
high-energy photons have the same helicities, $P_c =\xi_2=1$. The
dashed curves correspond to $\xi^2=0$. }
 \label{f3}
 \vspace{1.2cm}
\end{figure}

\begin{figure}[!t]
\includegraphics[width=0.47\textwidth]{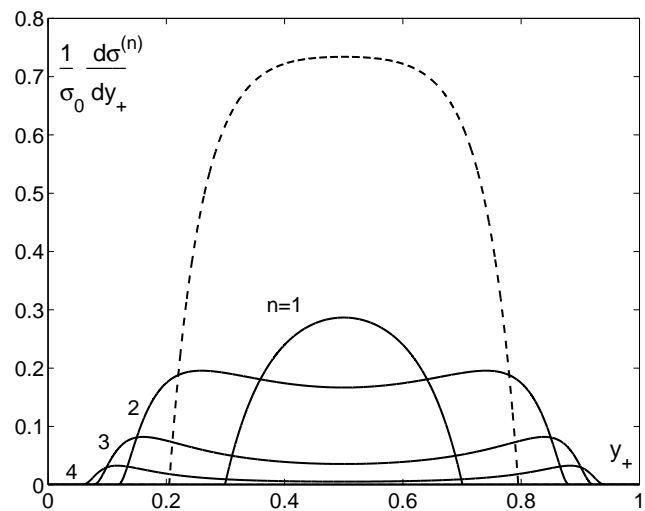}
 \caption{The same as in Fig. 3, but for $P_c =-\xi_2=1$.  }
 \label{f4}
\end{figure}

The following figures illustrate a dependence of the differential
cross sections $d \sigma^{(n)}(\mbox{\boldmath$\xi$},\,
\tilde{\mbox{\boldmath$\xi$}})/ dy_+$ and the positron
polarizations $\zeta_j^{(n)(f)+}$ on the relative positron energy
$y_+\approx {E_+/ \omega_1}$. The non-linearity parameter is
chosen the same as in the TESLA project, $\xi^2=0.3$, therefore,
the threshold value of $x$ for the first harmonic is
$x_1=4(1+\xi^2)=5.2$. We chose the value of parameter $x=6.2$,
which is a one unit above the threshold $x_1$.

{\it The case of the circularly polarized laser photons (Figs.
\ref{f3}, \ref{f4}, \ref{f5}, \ref{f6}).}

The spectra of the few first harmonics are shown in Figs. 3 and 4.
When helicities of the initial photons are the same ($\xi_2
P_c=1$, see Fig. 3), the main contribution (for the chosen
$\xi^2=0.3$) is given by the first harmonic, the probabilities for
generation of the higher harmonics are smaller. When helicities of
the initial photons are opposite ($\xi_2 P_c=-1$, see Fig. 4), the
first and the second harmonics give approximately equal
contributions. The changes in the spectra for the transition from
$\xi^2=0$ to $\xi^2=0.3$ are more pronounced for the case $\xi_2
P_c=-1$ than for $\xi_2 P_c=1$.

The polarization of positrons for these two cases are shown in
Fig. 5 and 6, respectively. It is seen that in the high-energy
part of the spectrum the longitudinal polarization of positrons
$\zeta_3^{(n)(f)+}$ is large for all harmonics and that its sign
coincides with the sign of the high-energy photon helicity
$\xi_2$.

\begin{figure}[!t]
\includegraphics[width=0.47\textwidth]{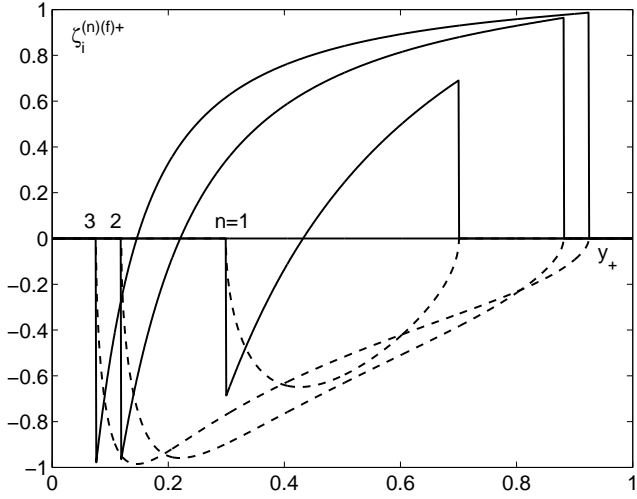}
 \caption{Polarization of positrons for different
harmonics $n$ at $x=6.2$ and $\xi^2=0.3$. The laser and
high-energy photons have the same helicity, $P_c =\xi_2=1$. The
solid curves correspond to $\zeta_3^{(n)(f)+}\approx 2\langle
\lambda_+ \rangle$, the dashed curves correspond to
$\zeta_2^{(n)(f)+}$ which is the transverse electron polarization
in the scattering plane; the transverse electron polarization
perpendicular to the scattering plane $\zeta_1^{(n)(f)+}=0$ in
this case.}
 \label{f5}
\end{figure}

\begin{figure}[!t]
\includegraphics[width=0.47\textwidth]{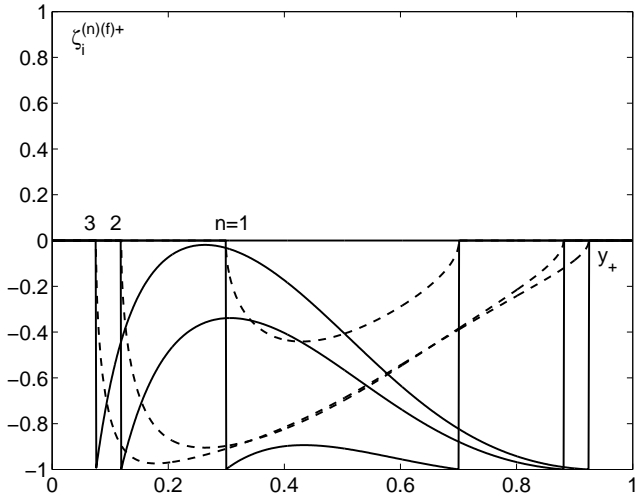}
 \caption{The same as in Fig. 5, but for $P_c =-\xi_2=1$. }
 \label{f6}
\end{figure}

{\it The case of the linearly polarized laser photons (Figs.
\ref{f7}, \ref{f8}).}

The spectra of the first few harmonics for this case are shown in
Fig. 7. They differ considerably from those for the case of the
circularly polarized laser photons shown in Fig. 3. First of all,
in the considered case the spectra depend on the polarization of
the high-energy photon only via the factor $\check{\xi}_3$ defined
in (\ref{81}). The maximum of the first harmonic at $y=y_1$ now is
about two times smaller than that on Fig. 3. Besides, the
harmonics with $n>1$ do not vanish at $y=y_n$ contrary to such
harmonics on Fig. 3.

\begin{figure}[!h]
\includegraphics[width=0.47\textwidth]{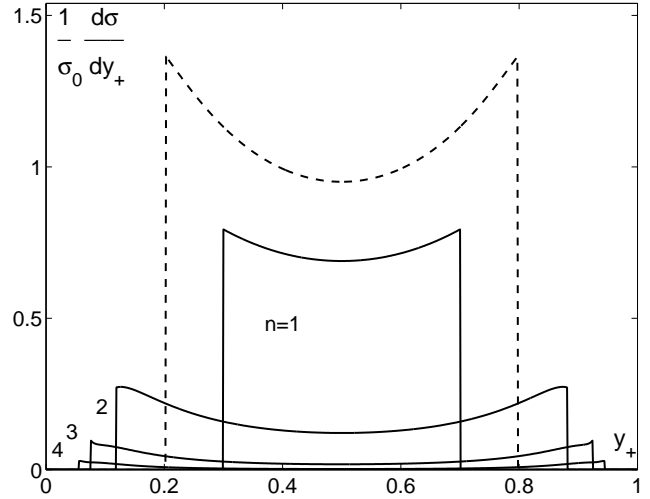}
 \caption{Energy spectra of positrons for different
harmonics $n$ at $x=6.2$ and $\xi^2=0.3$. The laser photons are
linearly polarized, the high-energy photon is circularly polarized
$\xi_2=1$. The dashed curves correspond to $\xi^2=0$. }
 \label{f7}
\end{figure}

\begin{figure}[!htb]
\includegraphics[width=0.47\textwidth]{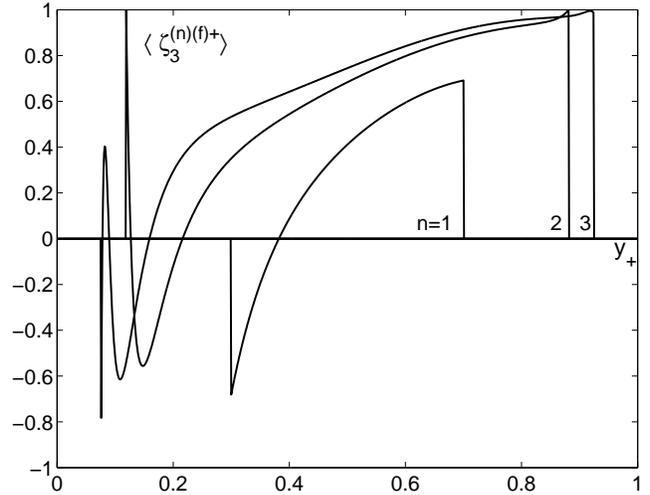}
\caption{The longitudinal polarization
$\langle\zeta_3^{(n)(f)+}\rangle = \langle G_3^{(n)+}\rangle /
 \langle \bar{F}_0^{(n)}\rangle$ of positrons for different
harmonics $n$ at $x=6.2$ and $\xi^2=0.3$, averaged over the
azimuthal angle $\varphi_+$. The laser photons are linearly
polarized, the high-energy photon is circularly polarized
$\xi_2=1$. The averaged transverse electron polarizations
$\langle\mbox{\boldmath$\zeta$}_{\perp}^{(n)(f)+}\rangle=0$ in
this case.}
 \label{f8}
\end{figure}

The polarization of positrons is not equal zero only if the
high-energy photon is circularly polarized. In this case the
longitudinal polarization of positrons is large in the high-energy
part of their spectrum (see Fig. 8), while the transverse positron
polarization, averaged over positrons azimuthal angle $\varphi_+$,
equals zero.

\section{Summary and comparison with other papers}

Our main results are given by (\ref{56})--(\ref{60}) for the
circularly and by (\ref{74})--(\ref{77}) for the linearly
polarized laser beam. They are expressed in terms of the 16
functions $\bar{F}_0,\; G_j^\pm$ and $\bar{H}_{ij}$ with $i,\,j =
1\div 3$, which describe completely the polarization properties of
the non-linear Breit-Wheeler process in a compact invariant form.
The function $\bar{F}_0$ enters the total cross section
(\ref{46}), (\ref{80}) and the differential cross sections
(\ref{79}), (\ref{80}). The polarization of the final positrons
(electrons) is described by functions $G_j^\pm$, which enter the
polarization vector $\mbox{\boldmath$\zeta$}^{(f)}_{\pm}$ given by
the exact equations (\ref{48}), (\ref{90}) and the approximate
equations (\ref{97}).

We considered the kinematics and the approximate formulae relevant
for the problem of $e\to \gamma$ conversion at the $\gamma \gamma$
and $\gamma e$ colliders. Besides, we discuss the spectra and the
polarization of the final particles in the limit of small and
large energies (Sect. 5). For the circularly polarized laser
photons and for large values of the parameter $x \gg 1+\xi^2$, we
obtain (with logarithmic accuracy) the simple analytical
expression for the total cross section, (\ref{110}), (\ref{111}).
In Sect. 6 we present the numerical results and discuss several
interesting features, which may be useful for the analysis of the
background and the luminosity of $\gamma e$ and $\gamma \gamma$
colliders.

Let us compare our results with those obtained earlier.

{\it Circular polarization of laser photons}. In the literature we
found  the results which can be compared with our functions
$\bar{F}_0$ and $G_3^\pm$, namely, in~\cite{NNR,R-review} (the
function $\bar{F}_0$), in~\cite{Tsai,Yokoya,GS} (the functions
$\bar{F}_0$ and $G_3^\pm$ only at
$\zeta_1^\pm=\zeta_2^\pm=\xi_1=\xi_3=0$).

The function $\bar{F}_0^{(n)}$ (\ref{56}) can be presented in the
form
 \be
\bar{F}^{(n)}_0= \langle \bar{F}^{(n)}_0 \rangle - C^{(n)}\xi_3\,,
 \label{a}
 \ee
where $\langle \bar{F}^{(n)}_0 \rangle$ is given by (\ref{82}) or
(using (\ref{81})) in the form
 \bea
\bar{F}^{(n)}_0&=&\langle \bar{F}^{(n)}_0 \rangle
 \ \label{b}\\
&&+ C^{(n)}\check{\xi}_3\cos{2\varphi_+} +C^{(n)}\check{\xi}_1
\sin{2\varphi_+}\,.
 \nn
 \eea
Our function $\langle \bar{F}^{(n)}_0 \rangle$ coincide with those
obtained in~\cite{NR,R-review,Yokoya,GS}, but our expression for
$C^{(n)}$,
 \bea
 C^{(n)}&=&2(f_n-g_n)+ s^2_n (1+\Delta)\,g_n
  \label{c}
  \\
  &=&
{4\over \xi^2}J_{n}^{2}(z_n)+ 4\left[ -\left( {n^2\over z_n^2}
-1\right)\, J_{n}^{2}(z_n)+\left(J_{n}^\prime(z_n)\right)^2\right]
 \nn
 \eea
does not coincides with that given by equations
(\ref{50})--(\ref{51}) on page 66 in~\cite{R-review}
 \bea
 C^{(n)}_{\rm R}&&
 \label{d}
  \\
  &&={4\over \xi^2}J_{n}^{2}(z_n)+ 4\left[ +\left( {n^2\over z_n^2}
-1\right)\,
J_{n}^{2}(z_n)+\left(J_{n}^\prime(z_n)\right)^2\right].
 \nn
 \eea
However, our expression has a proper limit: $C^{(1)}= s_1^2$ at
$\xi^2\to 0$, while the expression (\ref{d}) has a wrong limit
$C^{(1)}_{\rm R} = 2+s_1^2$ at $\xi^2\to 0$. The difference is due
to misprint in the paper~\cite{R-review}: the second term in
(\ref{d}) should have the sign ``minus''.

For polarization of the final electron, our results differ
slightly from those in~\cite{Yokoya}. Namely, function $G_3^-$ in
this paper coincide with our one. However, the polarization vector
$\mbox{\boldmath$\zeta$}_-^{(f)}$ in the collider system is
obtained in paper~\cite{Yokoya} not in the exact form (\ref{90}),
but only in the approximate form equivalent to our approximate
equation (\ref{97}).

{\it Linear polarization of laser photons}. In the literature we
found  the results which can be compared with our function
$\bar{F}_0$ only. Our expression for $\bar{F}_0^{(n)}$ (\ref{74})
is in agreement with that obtained in~\cite{NR,R-review}.

The correlation of the final particles' polarizations are
described by functions $\bar{H}_{ij}$ given in (\ref{60}) for the
circular, and in (\ref{77}) for the linear laser polarization. To
the best of our knowledge there are no any results in the
literature to compare with (\ref{60}) and (\ref{77}).

\section*{Acknowledgments}

We are grateful to I.~Ginzburg, M.~Galynskii and V.~Telnov for
useful discussions. This work is partly supported by RFBR (code
03-02-17734) and by INTAS; D.Yu.I. is supported by grant  DFG 436.

\section*{Appendix: Limit of the weak laser field}

Let us consider the limit of $\xi^2\to 0$, i.e. the process
(\ref{1}). The cross section of this process can be obtain from
the cross section (\ref{42}) at $\xi^2\to 0$ and has the form
 \bea
&&d\sigma(\mbox{\boldmath$\xi$},\,
\tilde{\mbox{\boldmath$\xi$}},\,\mbox{\boldmath$\zeta$}_\pm) =
{r_e^2\over 4x}\;\bar{F}\;d\Gamma\,,
 \label{A1}
 \\
&&d\Gamma = \delta (k_1+k_2-p_+-p_-)\;{d^3p_+\over
E_+}{d^3p_-\over E_-}\,,
 \nn
 \eea
where
 \bea
\bar{F}&=&\bar{F}_0+\sum ^3_{j=1}\left( G_j^{+}\zeta^+_j\; +
\;G_j^{-} \zeta^{-}_j\right)
 \nn\\
 && + \sum
^3_{i,j=1}\bar{H}_{ij}\,\zeta^{+}_i\,\zeta^{-}_j \,.
 \label{A2}
 \eea
To obtain $\bar{F}$ from $\bar{F}^{(1)}$ (\ref{45}), we should
take into account that the Stokes parameters of the laser photon
have the values (\ref{50}) for the circular polarization and the
values (\ref{65}) for the linear polarization. Besides, our
invariants $c_1$, $s_1$, $r_1$ and the auxiliary functions
(\ref{73}) transform at $\xi^2\to 0$ to
 \bea
c_1&\to& c=1- 2r\,,\;\; s_1\to s =2\sqrt{r(1-r)}\,,
 \nn
 \\ r_1&\to& r={u\over x}\,,\;\;u={1\over y_+ y_-}\,,
  \label{A3}
  \\
X_1 &\to& -c\,,\;\; Y_1 \to c(1-\tilde{\xi}_3)\,,\;\; V_1 \to
-\tilde{\xi}_3\,.
 \nn
 \eea
As a result, we obtain
 \bea
\bar{F}_0&=&(u-2)(1- c\,\xi_2 \tilde{\xi}_2)\,+
s^2(1-\xi_3)(1-\tilde{\xi}_3)
 \nn
 \\
&&-2(c\xi_1 \tilde{\xi}_1+\xi_3 \tilde{\xi}_3) \,;
  \label{A4}
 \eea
  \bea
G_1^{+}&=&-{s\over y_-}  \xi_1\, \tilde{\xi}_2+{s\over y_+}
\xi_2\, \tilde{\xi}_1 \,,
 \nn
 \\
G_2^{+}&=& {s \over y_+}\,\left[c(1-\tilde{\xi}_3)
\xi_2-\tilde{\xi}_2\right]+ {s\over y_-} \,\xi_3 \tilde{\xi}_2 \,,
 \label{A5}\\
G_3^{+}&=&(y_+ -y_-)u\,(\xi_2- c\,\tilde{\xi}_2) + {s^2\over
y_+}\, \xi_2(1-\tilde{\xi}_3)\,;
  \nn
 \eea
 \bea
\bar{H}_{11}&=& 2(1-c \xi_2 \tilde{\xi}_2)- (u-2)(c\xi_1
\tilde{\xi}_1+\xi_3 \tilde{\xi}_3)
 \nn\\
&&- s^2(1-\xi_3)(1-\tilde{\xi}_3)\,,
 \nn\\
\bar{H}_{22}&=& 2(1-c \xi_2 \tilde{\xi}_2)- (u-2)(c\xi_1
\tilde{\xi}_1+\xi_3 \tilde{\xi}_3)
 \nn\\
&& - s^2(1-\tilde{\xi}_3)[1+(u-1)\xi_3)\,,
 \nn\\
\bar{H}_{33}&=&-(u-2)(1-c \xi_2 \tilde{\xi}_2)+2(c\xi_1
\tilde{\xi}_1+\xi_3 \tilde{\xi}_3)
 \nn\\
&& + s^2(1-\tilde{\xi}_3)[u-1+\xi_3)\,,
 \label{A6}\\
\bar{H}_{12}&=& {s^2 \over y_-}\,\xi_1 (1-\tilde{\xi}_3)
 +u(y_+ -y_-) (\xi_1 \tilde{\xi}_3-c\xi_3 \tilde{\xi}_1)\,,
  \nn\\
\bar{H}_{13}&=& {s \over y_+}\,\tilde{\xi}_1
 - {s \over y_-}\, [\xi_3 \tilde{\xi}_1+c\xi_1(1- \tilde{\xi}_3)]\,,
 \nn \\
\bar{H}_{23}&=& {s \over y_-}[c\xi_3(1-\tilde{\xi}_3) -\xi_1
\tilde{\xi}_1]+{s \over y_+}[c(1-\tilde{\xi}_3) -\xi_2
\tilde{\xi}_2]\,.
  \nn
 \eea
The rest of the functions, $G_j^{-}$, $\bar{H}_{21}$,
$\bar{H}_{31}$ and $\bar{H}_{32}$, can be obtained from the above
equations using  the symmetry  relations (\ref{62}):
 \bea
G_j^{-}(y_+,\,y_-)&=&  G_j^{+}(y_-,\,y_+) \,,
 \nn\\
\bar{H}_{ij}(y_+,\,y_-)& =& \bar{H}_{ji}(y_-,\,y_+)\,.
 \label{A7}
 \eea



\begin{thebibliography}{99}
\bibitem{Yokoya}
K.~Yokoya, CAIN2e Users Manual. The latest version is available
from ftp://lcdev.kek.jp/pub/Yokoya/cain235/\\CainMan235.pdf.
\end{thebibliography}
\end{document}